\documentclass[a4paper,12pt]{amsart}
\usepackage{newcent}
\usepackage{fullpage}

\newtheorem{theorem}{Theorem}
\newtheorem{lemma}[theorem]{Lemma}
\newtheorem{observation}[theorem]{Observation}

\newtheorem{corollary}[theorem]{Corollary}

\def\paren#1{\ensuremath{\left({#1}\right)}}

\def\Real{\ensuremath{\mathbb{R}}}

\def\rank{\text{rank}}
\def\odd#1{\text{odd}\ensuremath{(#1)}}
\def\num#1{\ensuremath{{}^{\#}{#1}}}

\begin{document}
\title{Tight Bounds on the Complexity of Recognizing Odd-Ranked~Elements}
\author{Shripad Thite}
\curraddr{Department of Mathematics and Computer Science, Technische
Universiteit Eindhoven, Postbus~513, 5600~MB Eindhoven, The
Netherlands; Email:~sthite@win.tue.nl}
\date{May 23, 2006}
\maketitle

\begin{abstract}
Let $S = \langle s_1, s_2, s_3, ..., s_n \rangle$ be a given vector of
$n$ real numbers.  The \emph{rank} of $z \in \Real$ with respect to $S$ is defined as the number of elements $s_i \in S$ such that $s_i \le z$. We consider the following
decision problem: determine whether the odd-numbered elements $s_1$,
$s_3$, $s_5$, $\ldots$ are precisely the elements of $S$
whose rank with respect to $S$ is odd. We prove a bound of $\Theta(n \log n)$ on the number of operations required to solve this problem in
the algebraic computation tree model.
\end{abstract}

\bigskip

Let $S = \langle s_1, s_2, s_3, \ldots, s_n \rangle \in \Real^n$ be a given vector. For an arbitrary real $z$, define the \emph{rank} of $z$ with respect to $S$, denoted
by $\rank_S(z)$, as the number of elements of~$S$ less than or equal to
$z$. Thus, for instance, the largest element of $S$ has rank~$n$.
Let $\odd{S}$ denote the set of elements of~$S$
whose rank with respect to~$S$ is odd.

We consider the following problem: determine whether the odd-numbered elements $s_1$, $s_3$, $s_5$, $\ldots$ are precisely the elements of $S$ whose rank with respect to $S$ is odd. Without loss of generality, we can assume that $n$ is even because, otherwise, we can append an extra element $+\infty$ without changing the answer.

Note that $\odd{S}$ has size $n/2$ if and only if all $n$ values $s_i \in S$ are distinct; hence, the answer is `yes' only if $S$ is a vector of $n$ distinct numbers.

We prove matching upper and lower bounds on the number of operations required to solve
the problem in the algebraic computation tree (ACT) model (see
Ben-Or~\cite{ben-or83lowerbounds}).

The following algorithm solves the problem using $O(n \log n)$ comparisons. Sort $S' = \langle s_1, s_3, s_5, \ldots, s_{n-1} \rangle$ in non-decreasing order with an optimal sorting algorithm. Similarly, sort $S$ in non-decreasing order. Then, scan the vector $S'$ and the odd-numbered elements of $S$ to decide whether the two are equal.

Next, we prove the matching lower bound.

For a vector $S = \langle s_1, s_2, s_3, \ldots, s_n \rangle$, let $\sigma(S)$ denote the permuted vector $\langle s_{\sigma(1)},\allowbreak{} s_{\sigma(2)},\allowbreak{} s_{\sigma(3)},\allowbreak{} \ldots,\allowbreak{} s_{\sigma(n)} \rangle$. We call a permutation $\sigma$, where $\sigma(i)$ is odd if and only if $i$ is odd, a \emph{permissible} permutation.

\begin{lemma}
  There are $\paren{\paren{\frac{n}{2}}!}^2$ permissible
  permutations of a vector of $n$ elements.
\label{lemma:numpermissible}
\end{lemma}
\begin{proof}
There are $\frac{n}{2}!$ permutations of $n$ elements that permute the
$n/2$ odd-numbered elements only, and $\frac{n}{2}!$ that permute the
$n/2$ even-numbered elements only. A permissible permutation of $n$
elements is any composition of two permutations, one that permutes the
odd-numbered elements only and one that permutes the even-numbered
elements only.
\end{proof}

\begin{observation}
  A permutation $\sigma$ is permissible if and only if its inverse
  $\sigma^{-1}$ is permissible.
\label{obs:inverseispermissible}
\end{observation}

Let $W \subset \Real^n$ be the set of inputs for which the answer to
the question posed in the problem is `yes'. Recall that every point in
$W$ corresponds to a set of $n$ distinct real numbers.

\begin{lemma}
  For an arbitrary point $X \in W$, there is a unique permutation
  $\sigma$ that sorts $X$, i.e., such that $x_{\sigma(1)} <
  x_{\sigma(2)} < x_{\sigma(3)} < \ldots < x_{\sigma(n)}$. Moreover,
  such a permutation $\sigma$ is permissible.
\label{lemma:uniquesort}
\end{lemma}
\begin{proof}
The uniqueness of the sorting permutation $\sigma$ follows because
every point in $W$ corresponds to a set of distinct reals. When $X$ is
sorted, the odd-ranked elements must occupy the odd-numbered positions
of the sorted vector. Since $X \in W$, the odd-ranked elements are
already in odd-numbered positions of the original vector
$X$. Therefore, the permutation $\sigma$ is permissible.
\end{proof}

Let $\sigma_X$ denote the sorting permutation for $X$.

\begin{observation}
  If $\sigma_X$ is a permissible permutation, then $X \in W$.
\label{obs:sortispermissible}
\end{observation}

\begin{lemma}
  For every permissible permutation $\sigma$, there is a point $X \in
  W$ such that $\sigma = \sigma_X$.
\label{lemma:permissible}
\end{lemma}
\begin{proof}
Set $X = \langle \sigma^{-1}(1), \sigma^{-1}(2), \sigma^{-1}(3),
\ldots, \sigma^{-1}(n) \rangle$. We have,
\begin{align*}
\sigma(X) &= \langle \sigma(\sigma^{-1}(1)),
                     \sigma(\sigma^{-1}(2)),
                     \sigma(\sigma^{-1}(3)),
                     \ldots,
                     \sigma(\sigma^{-1}(n)) \rangle\\
          &= \langle 1, 2, 3, \ldots, n \rangle
\end{align*}
Therefore, $\sigma(X)$ is sorted, and by
Lemma~\ref{lemma:uniquesort}, it is the unique permutation that
sorts $X$; hence, $\sigma = \sigma_X$.

It remains to show that the point $X$ that we chose belongs to
$W$. The set of real numbers represented by $X$ is
$\{1,2,3,\ldots,n\}$. Since $\sigma$ is permissible, so is
$\sigma^{-1}$ by Observation~\ref{obs:inverseispermissible}; hence,
$\sigma^{-1}(i)$ is odd if and only if $i$ is odd.  Therefore, the
$i$th component of the vector $X$ is odd if and only if $i$ is
odd, which means that ${X \in W}$.
\end{proof}

\begin{lemma}
  For every two points $X, Y \in W$ such that $\sigma_X \ne \sigma_Y$,
  the two points $X$ and $Y$ lie in different connected components of
  $W$.
\label{lemma:diffcomponents}
\end{lemma}
\begin{proof}
Since $X, Y \in W$, both $\sigma_X$ and $\sigma_Y$ are permissible
permutations, by Lemma~\ref{lemma:uniquesort}.

For every point $A = \langle a_1, a_2, a_3, \ldots, a_n \rangle \in W$
such that
\[
     a_{\sigma_X(1)} < a_{\sigma_X(2)} < a_{\sigma_X(3)}
                     < \ldots          < a_{\sigma_X(n)}
\]
we have $\sigma_A = \sigma_X$. Since $\sigma_X$ is permissible, so is
$\sigma_A$; by Observation~\ref{obs:sortispermissible}, this implies that
$A \in W$. Additionally, $A$ is in the same connected component of $W$
as $X$ because every convex combination $B$ of $A$ and $X$ satisfies $\sigma_B = \sigma_X$.

On the other hand, since $\sigma_Y \ne \sigma_X$, there exists an $i$ in the range $1 \le i \le n-1$ such that $y_{\sigma_X(i)} \ge y_{\sigma_X(i+1)}$. Then, $X$ and $Y$ cannot be in the same connected component of $W$ because they are separated by the hyperplane $y_{\sigma_X(i)} = y_{\sigma_X(i+1)}$; every point $P$ on this hyperplane lies outside $W$ because it corresponds to an input where $\odd{P}$ has fewer than $n/2$ elements.

We have thus shown that the region $R_X$ where
\[
  R_X
=
  \{ \langle a_1, a_2, a_3, \ldots, a_n \rangle \in W :
     a_{\sigma_X(1)} < a_{\sigma_X(2)} < a_{\sigma_X(3)}
                     < \ldots          < a_{\sigma_X(n)} \}
\]
is a maximal connected component of $W$ containing $X$ ($R_X$ also happens to be convex); since $\sigma_Y \ne \sigma_X$, the region $R_X$ does not contain $Y$.
\end{proof}

\begin{theorem}
  The set $W$ has $\paren{\paren{\frac{n}{2}}!}^2$ connected
  components.
\label{thm:numcomponents}
\end{theorem}
\begin{proof}
The set $W$ can be partitioned such that each part corresponds to a
permissible permutation $\sigma$; by
Lemma~\ref{lemma:permissible}, $\sigma = \sigma_X$ for some $X \in
W$. By Lemma~\ref{lemma:numpermissible}, $W$ is partitioned into
$\paren{\paren{\frac{n}{2}}!}^2$ parts.  By
Lemma~\ref{lemma:diffcomponents}, every two distinct permissible
permutations $\sigma$ and $\sigma'$ correspond to two different
connected components of $W$, one consisting of all points $X \in W$
for which $\sigma_X = \sigma$ and the other consisting of all points
$Y \in W$ for which $\sigma_Y = \sigma'$.
\end{proof}

\begin{corollary}
  Every algebraic computation tree that decides the membership problem
  in $W$ must have depth $\Omega(n \log n)$.
\end{corollary}
\begin{proof}
Ben-Or~\cite{ben-or83lowerbounds} has proved that the
minimum height of an algebraic computation tree deciding membership in
$W$ is $\Omega(\log \num{W})$ where $\num{W}$ is the number of connected components of $W$. By Theorem~\ref{thm:numcomponents}, such a tree must have depth $\Omega(n \log n)$.
\end{proof}

\subsubsection*{Acknowledgments}
Thanks to Mark de Berg, Jeff Erickson, Sariel Har-Peled, and Jan Vahrenhold for fruitful discussions.



\begin{thebibliography}{9}

\bibitem{ben-or83lowerbounds}
\newblock
  ``Lower Bounds for Algebraic Computation Trees''.
\newblock
  Michael Ben-Or.
\newblock
  In \textsl{Proc.\@ ACM Symposium on Theory of Computing}, pp.~80--86, 1983.

\end{thebibliography}
\end{document}